\documentclass[5p,twocolumn]{elsarticle}

\usepackage{amsmath,amsfonts,amssymb}

\usepackage{graphicx,latexsym}
\usepackage[usenames,dvipsnames]{xcolor}
\usepackage[colorlinks=true, linkcolor=blue, citecolor=blue, urlcolor=blue, breaklinks=true,
            pdftitle={Thermoelectric Current and Coulomb-Blockade Plateaus in a Quantum Dot}, 
            pdfauthor={Kristinn Torfason}]{hyperref}
\usepackage[all]{hypcap}

\usepackage[protrusion=true, auto=true, spacing=true, kerning=true, tracking=true]{microtype}
\usepackage{float}

\makeatletter
  \newcommand{\raisemath}[1]{\mathpalette{\raisem@th{#1}}}
  \newcommand{\raisem@th}[3]{\raisebox{#1}{\(#2#3\)}}
\makeatother

\newcommand*{\ud}{\mathop{}\!\mathrm{d}}
\newcommand*{\ct}{\mathcal{T}}
\newcommand*{\tr}{\operatorname{Tr}}

\newcommand{\bluesolidcircle}{$\vcenter{\hbox{\protect\includegraphics{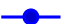}}}$}

\newcommand{\reddashedcircle}{$\vcenter{\hbox{\protect\includegraphics{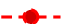}}}$}

\newcommand{\redsolidsquare}{$\vcenter{\hbox{\protect\includegraphics{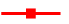}}}$}
\newcommand{\bluedashedcircle}{$\vcenter{\hbox{\protect\includegraphics{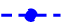}}}$}

\newcommand{\brownsoliddiamond}{$\vcenter{\hbox{\protect\includegraphics{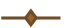}}}$}
\newcommand{\bluedashedsquare}{$\vcenter{\hbox{\protect\includegraphics{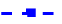}}}$}
\newcommand{\redsolidcircle}{$\vcenter{\hbox{\protect\includegraphics{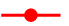}}}$}

\newcommand{\bluesolidsquare}{$\vcenter{\hbox{\protect\includegraphics{blue-solid-square.eps}}}$}
\newcommand{\brownsolidsquare}{$\vcenter{\hbox{\protect\includegraphics{brown-solid-square.eps}}}$}
\newcommand{\blackdottedtriangle}{$\vcenter{\hbox{\protect\includegraphics{black-dotted-triangle.eps}}}$}
\newcommand{\violetddashedcircle}{$\vcenter{\hbox{\protect\includegraphics{violet-ddashed-circle.eps}}}$}

\def\clap#1{\hbox to 0pt{\hss#1\hss}}

\journal{}

\makeatletter
\def\ps@pprintTitle{%
  \let\@oddhead\@empty
  \let\@evenhead\@empty
  \let\@oddfoot\@empty
  \let\@evenfoot\@oddfoot
}
\makeatother

\begin{document}

\begin{frontmatter}

\title{Thermoelectric Current and Coulomb-Blockade Plateaus in a Quantum Dot}

\author[hr,hi]{Kristinn Torfason}
\address[hr]{School of Science and Engineering, Reykjavik University, Menntavegi 1, IS-101 Reykjavik, Iceland}
\address[hi]{Science Institute, University of Iceland, Dunhaga 3, IS-107 Reykjavik, Iceland}

\author[hr]{Andrei Manolescu}
\author[hr]{Sigurdur I. Erlingsson}


\author[hi]{Vidar Gudmundsson}

\date{\today}

\begin{abstract}
A Generalized Master Equation (GME) is used to study the thermoelectric currents
through a quantum dot in both the transient and steady-state regime.
The two semi-infinite leads are kept at the same chemical potential but at
different temperatures to produce a thermoelectric current which has a varying sign
depending on the chemical potential. The Coulomb interaction between the electrons
in the sample is included via the exact diagonalization method. We observe a saw-teeth
like profile of the current alternating with plateaus of almost zero current. 
Our calculations go beyond the linear response with respect to the temperature gradient,
but are compatible with known results for the thermopower in the linear response regime.
\end{abstract}


\end{frontmatter}

\section{Introduction}

The electrical conduction of open nanoelectronic devices driven by
electric potentials or fields generated in various ways is a major
topic in mesoscopic physics.  Outside this area complementary research
on thermoelectric currents, thermopower, and related thermal properties
in the quantum regime for systems like quantum dots has been more difficult,
but on a growing trend in the last two decades.  Temperature
control down to the milli Kelvin range and temperature gradients at nanoscale are
attainable in the laboratories and generate new scientific opportunities~\cite{Giazotto}.

\begin{figure}[ht]
  \includegraphics[width=1.00\linewidth]{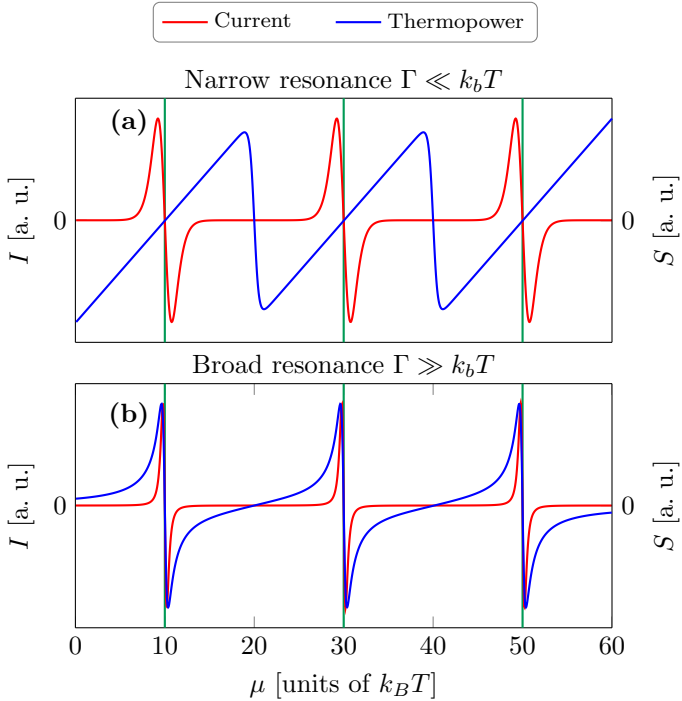}
  \caption{A qualitative behavior of the thermopower $S$ and thermoelectric current
$I$ generated as a linear response to a temperature bias between leads, for a quantum dot, 
as function of the chemical potential 
in the leads for two regimes: high temperature, such that the resonance width $\Gamma << k_BT$ 
(upper panel) and low temperatures, i.\ e.  $\Gamma >> k_BT$ . The red lines show the 
current and the green line shows the thermopower.  Saw teeth alternating with zero plateaus 
are obtained for the current.  The thermopower follows this behavior only at low temperatures.  
The chemical potential is in units of $k_BT$, but arbitrary units (a.u.) 
are used on the vertical axes. 
}
\label{fig:saw-tooth}
\end{figure}

The thermopower of quantum dots was initially studied theoretically by 
Beenakker and Staring~\cite{Beenakker}.  They calculated the Seebeck coefficient
\begin{equation}
S=-\lim_{\Delta T \to 0}\frac{V}{\Delta T} \ ,
\end{equation}
where $V$ is the voltage generated across the quantum dot weakly
connected to electron reservoirs at a temperature bias $\Delta T$,
under the condition that the current between the two reservoirs is zero.
They obtained oscillations of $S$ around $S=0$ as function of the Fermi energy
in the reservoirs, with symmetric positive and negative values. 
The Coulomb electron-electron interaction in the 
quantum dot was included in the charging (``orthodox") model, and the result
of it was a saw-tooth profile of the thermopower, with oscillations having
the positive slope smaller than the negative slope.  
The predicted results were confirmed in subsequent
experimental work by the same team~\cite{Staring} and also by
Dzurak et~al.~\cite{DzurakSSC}. Few years later Dzurak et~al. published
a new series of measurements which show that at temperatures below 100 mK the 
saw-tooth oscillations of the thermopower vs. the Fermi energy alternate
with plateaus of zero thermopower~\cite{DzurakPRB}. A qualitative graph of the 
thermopower is shown in Fig.~\ref{fig:saw-tooth}.

While the saw-teeth were attributed to sequential tunneling and high
temperatures~\cite{Beenakker}, the zero plateaus were initially attributed
to many-body effects.  Later on Turek and Matveev derived a theory of the
thermopower of quantum dots in which the zero plateau at low temperature
is a result of cotunneling~\cite{Turek}.  A more complex cotunneling
theory, beyond the limit of weak tunnel coupling, and including quantum fluctuations, was proposed
by Kubala and K\"onig~\cite{Kubala}, and later by Billings et~al. who
also included exchange effects~\cite{Billings}.  Further experimental
results were interpreted in terms of sequential-tunneling dominated
thermopower at high temperatures, leading to a saw-tooth profile, and cotunneling
onset at low temperatures, leading to zero plateaus~\cite{Scheibner}.
In a recent experimental paper by Svensson et~al.~\cite{Svensson} the thermopower 
of quantum dots is systematically investigated and the lineshape is
carefully analyzed in various conditions.  Periodic sequences of
a negative peak followed by positive peak followed by a zero plateau 
of the thermopower of the quantum dot as function of chemical potential in the 
leads are clearly seen over large intervals of chemical potentials.  
The interpretation of these results is done 
using a Landauer formula with an empirical transmission function.

In the constant interaction model the effects of the Coulomb interaction
are taken into account using only a charging energy, with no other effects
on the energy spectrum \cite{vanHouten92,Kouwenhoven01}.  The transport is
described by a series of resonances, and in between the resonance peaks
there are zero current plateaus.  This is the standard Coulomb blockade
picture \cite{Kouwenhoven01}.  In the Landauer-B\"uttiker approach this
phenomena is described using a transmission function $\mathcal{T}(E)$
representing the resonances in the transport, usually a Lorentzian of
width $\Gamma$ centered on some resonance energies $E_n$, and the current
is then given by \cite{BlanterButtiker}
\begin{eqnarray}
I(\mu)&=&\frac{2e}{h}\int dE \mathcal{T}(E) (f(E,T_L)-f(E,T_R)) \\
&\approx&-\frac{2e}{h}\frac{\Delta T}{T}\int dE \mathcal{T}(E) \frac{\partial f}{\partial E} (E-\mu),
\label{eq:LB}
\end{eqnarray}
where $T_L=T-\frac{\Delta T}{2}$ and $T_R=T+\frac{\Delta T}{2}$ are
the temperatures of the two contacts (left and right) having identical 
chemical potentials $\mu$.  There are two regimes that we want to consider 
from Eq.\ (\ref{eq:LB}): ($i$) $\Gamma \ll kT$ and ($ii$) $\Gamma \gg kT$.  In case ($i$)
the current is non-zero around resonance peaks in an interval $\sim kT$.  Between those intervals the current is exponentially suppressed
due to the derivative of the Fermi function.  In the other case, ($ii$), the current is proportional to the derivative of the
Lorentzian peak, resulting in an $1/E^3$ suppression of the current between peaks.
We assume both $\Gamma$ and $k_BT$ smaller than the energy separation
between resonances.

The temperature driven current given in Eq.\ (\ref{eq:LB}) will give rise to a voltage drop over the sample.  In linear response
the thermovoltage is given by
\begin{eqnarray}
V_\mathrm{Th}(\mu)&=&\frac{I(\mu)}{G(\mu)},
\end{eqnarray}
where $G$ is the conductance of the system.  The conductance in case ($i$) is determined by peaks of width $\sim kT$, with exponential suppression
between adjacent resonances (assuming $kT\ll \Delta E$).  The ratio of $I$ and $G$ will thus lead to a saw-tooth like pattern without plateaus in $V_\mathrm{Th}$,
even though there are plateaus in $I$.  In case ($ii$) the conductance is proportional to the Lorentzian peak, which tends to zero like $1/E^2$ outside
the resonances.  In this case the thermal voltage is saw-tooth like but with plateaus between resonances.

In the present paper we use the generalized master equation (GME) as a tool to understand
the electric currents generated in an open quantum dot due to a temperature bias.  The
dot is connected to external leads seen as electron reservoirs and kept at the same
chemical potential.  We obtain the currents in the leads produced by a 
finite temperature bias.  We calculate the time dependent 
currents when the leads are gradually coupled to the sample and we find numerically the
asymptotic currents in the leads in the steady state.
To our knowledge the master equation has not been commonly used for the 
thermoelectric response of open systems.  One approach was performed by Koch et~al.  
who used in fact rate equations, neglecting the off-diagonal elements
of the reduced density operator, but including cotunneling~\cite{Koch}.
Our method is not restricted to the linear response to a small temperature
gradient.  We obtain a line shape of the currents vs. the chemical potential
as illustrated in Fig.~\ref{fig:saw-tooth} for low temperatures, where 
zero plateaus are also expected for the thermopower~\cite{DzurakPRB, Scheibner, Svensson},
as illustrated in Fig.~\ref{fig:saw-tooth}.  
Our method however does not include cotunneling effects, but only sequential tunneling.
Instead, the Coulomb interaction in the dot is completely incorporated using the 
method of exact diagonalization.   We also discuss time dependent and transient currents
in the system and the effect of a third terminal attached to the
quantum dot.  The third terminal was proposed in order to create a phase-breaking
mechanism inside the dot~\cite{Sanchez}.

The paper is organized as follows: The model and the methodology are
described in Section~\ref{sec:model}, analytical calculations in Section~\ref{sec:gmeone},
the numerical results are presented in Section~\ref{sec:results},
and the conclusions in Section~\ref{sec:conclusion}.

\section{The Model\label{sec:model}}
\begin{figure}
    \centering
    \includegraphics[]{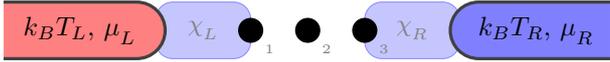}
    \caption{ A sketch of the system under study. A 1D lattice with 3~sites 
    (``the sample'') is connected to two semi-infinite leads via tunneling.
    The left lead is connected to the left end of lattice, while right lead
    is connected to the right end.
    The contacts (\(\chi_L\), \(\chi_R\)) are time-dependent.\label{fig:GME-system}}
  \end{figure}

The physical system consists of a sample connected to two leads acting
as particle reservoirs.  We shall adopt a tight-binding description of
the system: the sample is a short quantum wire and the leads are 1D and
semi-infinite. The sample can also be seen as an elongated quantum dot. 
In this work we consider a sample of 3~sites. This number
optimizes the computational time and the physical phenomenology which we
intend to describe.  A sketch is given in Fig.~\ref{fig:GME-system}.
The left lead (or the source, marked as~\textcolor{red}{$L$}) is
contacted at one end of the sample and the right lead (or the drain,
marked as~\textcolor{blue}{$R$}) is contacted at the other end. The
Hamiltonian of the coupled system reads as
  \begin{equation}\label{eq:total_ham}
    H(t) = \sum_{\ell} H_{\ell=L,R} + H_S + H_T(t) = H_0 + H_T(t)\, ,
  \end{equation}
where $H_S$ is the Hamiltonian of the isolated sample, including the
electron-electron interaction,
  \begin{equation}\label{eq:sample_ham}
    H_S = \sum_n E_n d_n^\dagger d_n + \frac{1}{2} \sum_{\substack{mn\\m'n'}} 
V_{mn,m'n'} d_m^\dagger d_n^\dagger d_{m'} d_{n'}\, .  
  \end{equation}
The (non-interacting) single-particle basis states have wave functions
$\{\phi_n\}$ and discrete energies $E_n$. $H_{\ell}$, with
${\{\ell\} = (L, R)}$, is the Hamiltonian corresponding to the left
and the right leads. The last term in Eq.~\eqref{eq:total_ham}, $H_T$
describes the time-dependent coupling between the single-particle basis
states of the isolated sample and the states $\{ {\psi}_{q\ell}\}$
of the leads:
  \begin{equation}\label{Htunnel}
    H_T(t)=\sum_{n}\sum_{\ell}\!\int\! \ud q\:\chi_{\ell}(t)(T^{\ell}_{qn}c^{\dagger}_{q\ell}d_n + \text{h.c.}) \,.
  \end{equation}
The function $\chi_{\ell}(t)$ describes the time-dependent switching
of the sample-lead contacts, while $d^{\dagger}_n$ and $ c_{q\ell}$
create/annihilate electrons in the corresponding single-particle
states of the sample or leads, respectively. The coupling coefficient
  \begin{equation}\label{eq:trans-coeff}
    T^{\ell}_{qn}=V_\ell{\psi}^{*}_{q\ell}(0)\phi_n(i_\ell)\, ,
  \end{equation}
involves the two eigenfunctions evaluated at the contact sites
$(0,i_\ell)$, $0$ being the site of the lead $\ell$, and $i_\ell$ the site
in the sample. The wave functions in the leads are 
${\psi_{q\ell}(0)=\sqrt{\sin q /2\tau}}$ with $\tau$ the 
hopping energy in the leads (the same for all leads), the energy spectrum of the 
leads being $\epsilon_{\ell}(q)=2\tau \cos q$  \cite{1367-2630-11-7-073019}. 
The hopping energy in the sample, denoted as $t_s$, will be considered different
than in the leads, and will be used 
as the energy unit.
In Fig.~\ref{fig:GME-system} the left lead is connected to the site
$i_L=1$ and the right lead on the site $i_R=3$.
The parameter $V_\ell$ plays the role of a coupling constant between the 
sample and the leads. 
%

We will ignore the Coulomb effects in the leads, where we assume a high
concentration of electrons and thus strong screening and fast particle
rearrangements. The GME is formulated in the Fock space and therefore it is natural to include
the Coulomb electron-electron interaction in the dot in a complete many-body
manner and to calculate the exact many-body states using a straight forward 
diagonalization in the basis of occupation numbers. This method is known as exact
diagonalization in the community of mesoscopic physics, but often called configuration 
interaction by quantum chemists.
The many-electron states (MES) are calculated in the Fock space built on
non-interacting single-particle states~\cite{PRBC}.  Since the sample is
open the number of electrons is not fixed, but the Coulomb interaction
conserves the number of electrons, which means the interacting eigenvectors are
linear combinations of the non-interacting eigenvectors with a fixed 
number of particles. 


The equation of motion for our system is the quantum Liouville equation,
  \begin{equation}
   i\hbar \dot{W}(t) = \left[ H(t), W(t) \right] \ ,  
  \end{equation}
where $W(t)$ is the statistical operator of the total system made by the
sample plus leads, which are connected at time $t=0$.  Before the
connection, at $t<0$, the sample and the leads are independent and in
equilibrium, meaning that $W(t < 0) = \rho_L \rho_R \rho_S$, i.~e. simply the
product of the density operator of the sample, $\rho_S$, the left lead
$\rho_L$ and the right lead $\rho_L$.

Following the Nakajima-Zwanzig technique~\cite{PhysRevB.77.195416}
we define the reduced density operator (RDO), $\rho(t)$, by tracing
out the degrees of freedom of the environment, the leads in our case,
over the statistical operator of the entire system, $W(t)$
  \begin{equation}\label{eq:rdo}
    \rho(t) = \tr_L\tr_R W(t)\, .
  \end{equation}
%
For a sufficiently weak coupling strength ($V_\ell$) one obtains
the non-Markovian integro-differential master equation for the RDO
\begin{subequations}\label{eq:gme-all}
  \begin{equation}\begin{split}\label{eq:gme}
    {\dot\rho}(t) = &-\frac{i}{\hbar}[H_S,\rho(t)]\\
                  &-\frac{1}{\hbar^2}\sum_{\ell}\!\int\! \ud q\:\chi_{\ell}(t)
      \Big(\left[{\cal T}_{q\ell},\Omega_{q\ell}(t)\right]+ \text{h.c.} \Big) \, ,
  \end{split}\end{equation}
where the operators $\Omega_{q\ell}$ and $\Pi_{q\ell}$ are defined as
  \begin{equation}\begin{split}\label{eq:gme-omega-pi}
    \Omega_{q\ell}(t)&=e^{-itH_S}\! \int_{0}^t \!\!\ud s\:\chi_{\ell}(s)\Pi_{q\ell}(s)e^{i(s-t)
    \varepsilon_{q\ell}}e^{itH_S} \, ,\\
    \Pi_{q\ell}(s)&=e^{isH_S}\left ({\cal T}_{q\ell}^{\dagger}\rho(s)(1-f_\ell)-\rho(s)
    {\cal T}_{q\ell}^{\dagger}f_\ell\right )e^{-isH_S} \, ,
  \end{split}\end{equation}
\end{subequations}
and \(f_\ell\) is the Fermi function of the lead \(\ell\) describing the state
of the lead before being coupled to the sample. The operators
\({\cal T}_{q\ell}\) and \({\cal T}_{q\ell}^{\dagger}\) describe the
'transitions' between two many-electron states (MES)
when one electron enters the sample or leaves it.

The GME is solved numerically by calculating the matrix elements of the
RDO in the basis of the interacting MES, in small time steps, following a
Crank-Nicolson algorithm. More details of the derivation of the GME can
be found in Ref.~\cite{1367-2630-11-7-073019}.  The calculation of 
the interacting MES is described in Ref.~\cite{PRBC}.
The switching functions $\chi_{\ell}(t)$ must be defined.  For example
any function starting at zero and gradually increasing to one can be 
used to obtain the steady state in the asymptotic limit.  In principle 
any other time dependence can be used, like steps or periodic functions.

Mean values of observables can by obtained by taking the trace of product of 
the corresponding operator and the RDO.
The total time dependent charge in the sample is found
by using the number operator \({{\cal N} = \sum_m d_m^\dagger d_m}\):
  \begin{equation}\label{eq:obs-chg}
    \langle Q(t) \rangle = e \tr\{\rho {\cal N}\} = e \sum_N N \sum_{\alpha_N} 
\langle \alpha_N | \rho(t) | \alpha_N \rangle\, ,
  \end{equation}
where \(\alpha_N\) denotes the (Coulomb interacting) MESs with fixed number of electrons \(N\).
Remark that one can also calculate the partial charge accumulated on \(N\)-particle MESs.

The currents in the leads are then found by taking the derivative of 
Eq.~\eqref{eq:obs-chg} with respect to time.
The time derivative of the RDO can be substituted by the right-hand side of the 
GME [Eq.~\eqref{eq:gme}] and so it is possible identify the currents in each lead,
  \begin{equation}\begin{split}\label{eq:cur-gme-rho}
    \langle J_\ell(t) \rangle & = -\frac{1}{\hbar^2} \sum_N N \sum_{\alpha_N} 
        \int\! \ud q\, \chi_\ell(t) 
\langle \alpha_N | \left[{\cal T}_{q\ell},\Omega_{q\ell}(t)\right] | 
\alpha_N \rangle \\
& + \text{h.c.}
  \end{split}\end{equation}
%

\section{GME With One Energy Level\label{sec:gmeone}}
  To better understand the thermal effects in the GME we solve the GME for a
sample with one single site (a single level quantum dot).
  To simplify Eq.~\eqref{eq:gme} we start by inserting
  \begin{equation}\begin{split}
    \rho(s) &= U_S^\dagger(t-s) \rho(t) U_S (t-s)\\
                     &= e^{iH_S(t-s)/\hbar} \rho(t) e^{-iH_S(t-s)/\hbar}\, ,
  \end{split}\end{equation}
  i.e. we propagate the density matrix backwards in time and take it outside the time integral in
  Eq.~\eqref{eq:gme-omega-pi}.  In fact this is the Markov approximation.

  The Fock space contains now only two states, the vacuum state with energy $0$ and a 
single particle states with energy $E > 0$.  We use here $E=2$ units of $t_s$. 
Of course, for a one-site model the hopping energy $t_s$ has no meaning, the 
result being only the diagonal term in the lattice Hamiltonian.
  We want to find the occupation of these states, i.~e. the diagonal elements of the RDO,
for this two-level model.
  In this case Eqs.\eqref{eq:gme-all} are greatly simplified because only one of the transfer matrix elements (\({\cal T}_{q\ell}\)) is non-zero:
  \begin{equation}
    \langle 1 | \ct_{q\ell} | 0 \rangle = \sum_n T_{qn}^\ell \langle 1 | d_n^\dagger | 0 \rangle = T_{q 1}^\ell\, .
  \end{equation}
  This simplifies the commutator Eq.~\eqref{eq:gme} and gives
  \begin{equation}
   \frac{\partial}{\partial t} \langle 0 | \rho | 0 \rangle = 
     -\frac{2}{\hbar^2}\sum_\ell \int \ud q \operatorname{Re}\Big\{ \langle 0 | \Omega_{q\ell} | 1 \rangle \langle 1 | \ct_{q\ell} | 0 \rangle \Big\}\, .
\label{eq:gme-simple}
  \end{equation}
  The trace of the RDO is 1, therefore the other diagonal element is given by \(\langle 1 | \rho | 1 \rangle = 1 - \langle 0 | \rho | 0 \rangle\).
  Only one matrix element for \(\Omega\) is needed to find the diagonal elements of the RDO
  \begin{equation}\begin{split}
    \langle 0 | \Omega_{q\ell} | 1 \rangle
     &= \frac{\langle 0 | \ct^\dagger_{q\ell} | 1 \rangle}{\epsilon_\ell(q) - E} 
     i\left\{ e^{-it(\epsilon_\ell(q) - E)} - 1\right\}\\
     \times&\Big( \langle 1 | \rho | 1 \rangle (\, 1 - f_\ell\big(\epsilon_\ell(q)\big)\,) - \langle 0 | \rho | 0 \rangle f_\ell\big(\epsilon_\ell(q)\big) \Big)\, .
  \end{split}\end{equation}
  The \(q\)-integral can be evaluated in the steady-state limit if we admit that 
  for \({t \rightarrow \infty}\)
the real part of \(\langle 0 | \Omega_{q\ell} | 1 \rangle\) contains a delta function
  \(\delta\big(\epsilon_\ell(q) - E\big)\).  With this approximation we actually neglect the 
level broadening $\Gamma$ due to the lead-dot coupling \cite{Li}.  That broadening will be 
discussed in the next section. 
  In the steady-state limit the diagonal values of \(\rho\)  approach a constant value and
thus the right hand side of Eq.\ (\ref{eq:gme-simple}) must be zero. Using Eq.\ (\ref{eq:trans-coeff}) and the 
delta function one obtains
\begin{equation}
\sum_{\ell} V_{\ell}^2\left[\langle 1|\rho|1 \rangle \left(1-f_\ell(E)\right) - \langle 0|\rho| \rangle f_\ell(E)\right] =0 \ .
\end{equation}
  Therefore the occupations in the steady-state are
  \begin{equation}\label{eq:chg-ana}
    \langle 0 | \rho | 0 \rangle = 1 - \frac{\sum_\ell V_\ell^2 f_\ell(E)}{\sum_\ell V_\ell^2}\, ,\quad \langle 1 | \rho | 1 \rangle = 1 - \langle 0 | \rho | 0 \rangle\, ,
  \end{equation}
  and the current is
  \begin{equation}\label{eq:cur-ana}
   J_\ell = \frac{V_\ell^2}{\tau^2} \Big\{ \langle 0 | \rho | 0 \rangle f_\ell(E) - \langle 1 | \rho | 1 \rangle \big(1-f_\ell(E)\big) \Big\}\, .
  \end{equation}
The equation for the occupation and the
  currents in the leads was derived using the fact that in the steady-state \(\rho\) is constant and thus its time derivative is zero.
  This implies that in the steady state the currents in the left and right leads are equal: what goes into the sample must also go out. Therefore using Eq.~\eqref{eq:chg-ana} we can eliminate \(\rho\) from
  Eq.~\eqref{eq:cur-ana} and we obtain:
  \begin{equation}\begin{split}
    J_L = \frac{1}{\tau^2} \frac{V_L^2V_R^2}{V_L^2 + V_R^2} \Big( f_L(E) - f_R(E) \Big)\, ,\\
    J_R = \frac{1}{\tau^2} \frac{V_L^2V_R^2}{V_L^2 + V_R^2} \Big( f_R(E) - f_L(E) \Big)\, .
  \end{split}
\label{eq:JLR}
\end{equation}
  Which makes it clear that in the steady-state the currents are equal, but with opposite signs $J_L=-J_R$, with positive ``in" from the left led, and negative ``out" into the right lead.
  \begin{figure}[htb]
    \includegraphics[width=1.0\linewidth]{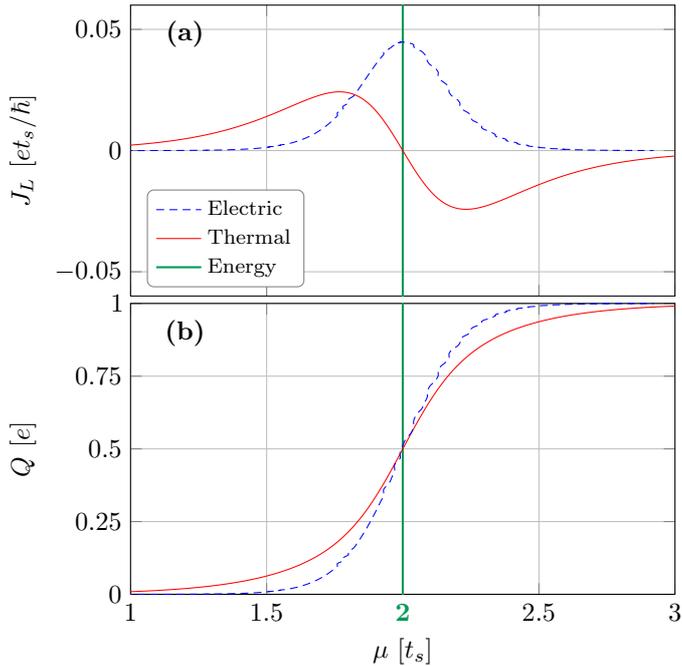}
    \caption{ Analytical calculations for a system with one site. \textbf{(a)} The red solid line 
             shows the currents in the left lead for a thermal bias, but no electrochemical bias.  
             The blue dashed line shows the current for an electrochemical bias, but no thermal bias. 
             The thermal bias used in the calculations was~\({k_BT_L = 0.25}\),~\({k_BT_R = 0.10}\), 
             with $\mu=\mu_L=\mu_R$.  The electrochemical bias was~\({\Delta \mu = 0.15}\), 
             with \({k_BT_L = k_BT_R = 0.10}\).
             \textbf{(b)}~Charge in the system in the two cases.
             On the  horizontal axis we use the chemical potential $\mu=\mu_L=\mu_R$ for the thermal bias and the  mean value 
          \({\mu = (\mu_L + \mu_R)/2}\) for the electrochemical bias. The green vertical line shows the energy of the 
           resonance, $E=2$.
         \label{fig:num-cur}}
  \end{figure}
The results of the analytical calculations
can be seen in Fig.~\ref{fig:num-cur}. In
Fig.~\hyperref[fig:num-cur]{\ref{fig:num-cur}(a)} we show the
results for a thermal bias, i.~e. for different temperatures
in the leads, $T_L>T_R$ and $\mu_L = \mu_R$, and also for an
electrochemical bias, i.~e.  $T_L=T_R$ and $\mu_L > \mu_R$.
Fig.~\hyperref[fig:num-cur]{\ref{fig:num-cur}(b)} shows the charge
in the system calculated using Eq.~\eqref{eq:chg-ana}.  The current
generated thermically can be positive or negative, the transition
occurring when the Fermi level is equal to the energy of the single
level~\(\mu_{\raisemath{-2pt}{L,R}} = E\).  At this point the state is
half filled and the current in the leads is zero. The probability of
a transition to/from the energy level to/from the left and right lead
is equal. When the state is less than half filled a transition from the
energy level to the right lead has a higher probability than to the left
lead. This gives a positive current flowing from the left to the right.
Once the state is more than half filled a transition from it to the left
lead is more probable. Giving a negative current flow from the right to
the left. This difference in transition probability is due to different
temperatures at the left and right lead.

According to Eq.\ (\ref{eq:JLR}) the current is essentially the difference
of two Fermi functions.  For the thermoelectric current they are centered
at the same chemical potential, but they have different widths.  Tor the pure
electric current they are centered at two different chemical potentials,
but they have the same width.  Therefore the width of the current peaks in
Fig.\ \ref{fig:num-cur} are only due to the temperatures or chemical
potentials.  For example the width of the current created by the chemical
potential bias, at half height, should be $\Delta\mu + 2k_BT=0.15+2\times
0.1=0.35$, which is consistent with Fig.\ \ref{fig:num-cur}. 
(The level width $\Gamma$ has been neglected.)

We can conclude this section with the idea that the thermoelectric currents can be
understood as being related to the difference between two Fermi functions with the
same center but different widths.  Therefore, as pointed out recently by 
Tagani and Soleimani~\cite{Tagani}, there are two reasons for the current to become
zero:~(1) half filling, where the two Fermi functions are equal to~0.5, and (2)~integer
filling, where both are~0 or~1. This idea will help us to understand more complex  
results incorporating many-body effects.  For comparison the currents generated by an 
electrochemical potential bias are given by the difference between two Fermi functions
of the same width but with different centers, and thus have a constant sign and a maximum 
at half filling.

\section{Results for a Many-Body System\label{sec:results}}
  \begin{figure}[htb]
    \center
    \includegraphics[]{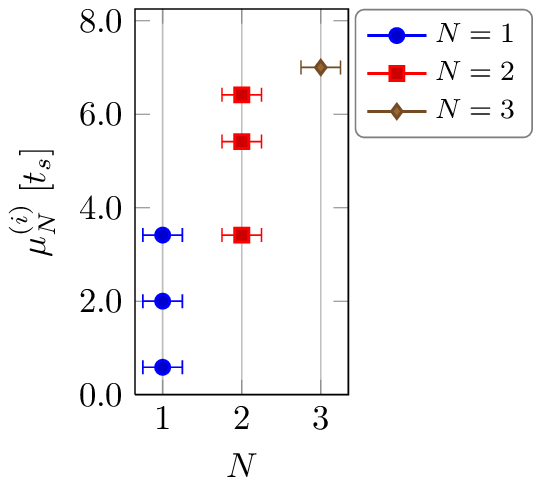}
    \caption{ Electrochemical potential diagram for a system with 3~sites and Coulomb interaction
             strength~\({u_c = 2.0}\). Single particle states blue dots~(\bluesolidcircle).
             Two particle states red squares~(\redsolidsquare).
             Three particle state brown diamond~(\brownsoliddiamond).\label{fig:mu-diag}}
  \end{figure}
The MESs of the sample are characterized by the electrochemical potentials
\({\mu_N^{(i)} := {\cal E}_N^{(i)} - {\cal E}_{N-1}^{(0)}}\), where
\({\cal E}_{N}^{(i)}\) is an energy of the sample spectrum containing \(N\)
particles, \({i=0}\) indicating the ground state and \({i>0}\) the excited
states. In Fig.~\ref{fig:mu-diag} we see the electrochemical potential diagram for a
system with three~lattice points and Coulomb interaction strength~\({u_c = 2.0}\)
(units of $t_s$).
The system has three single particle states with energies \({E_1^0 = 2 - \sqrt(2) \approx 0.59}\), \({E_1^1=2}\),
\({E_1^2 = 2 + \sqrt{2} \approx 3.41}\), three two-particle states with energies \({E_2^0=4}\), \({E_2^1=6}\),
\({E_2^2=7}\) and one three-particle state with energy \({E_3^0=11}\).
For the single-particle states~(\({N=1}\)) the chemical potentials are in fact the single-particle energies.
The effects of the Coulomb interaction on the two~(\({N=2}\)) and three-particle states~(\({N=3}\)) is to shift them
upwards. 

\begin{figure}[h]
    \includegraphics[width=1.0\linewidth]{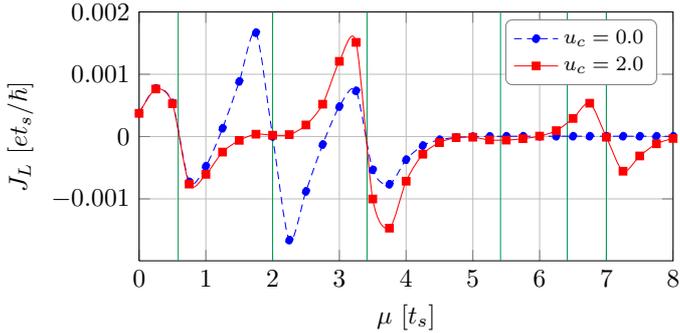}
    \caption{ Comparison of the current in the left lead with and without Coulomb interaction.
             Red solid line~(\redsolidsquare) has the Coulomb strength~\(u_c = 2.0\) while the blue dashed line~(\bluedashedcircle) has~\(u_c = 0.0\).
             The temperature is~\(k_BT_L = 0.25\) in the left lead and~\(k_BT_R = 0.10\) in the right lead.
             The lead-dot coupling parameters are $V_{\ell}=V_L=V_R=0.75$ and the hopping 
energy in the leads is $\tau=4$ (units of $t_s$).
  \label{fig:coul}}
  \end{figure}

\begin{figure}[H]
    \includegraphics[width=0.93\linewidth]{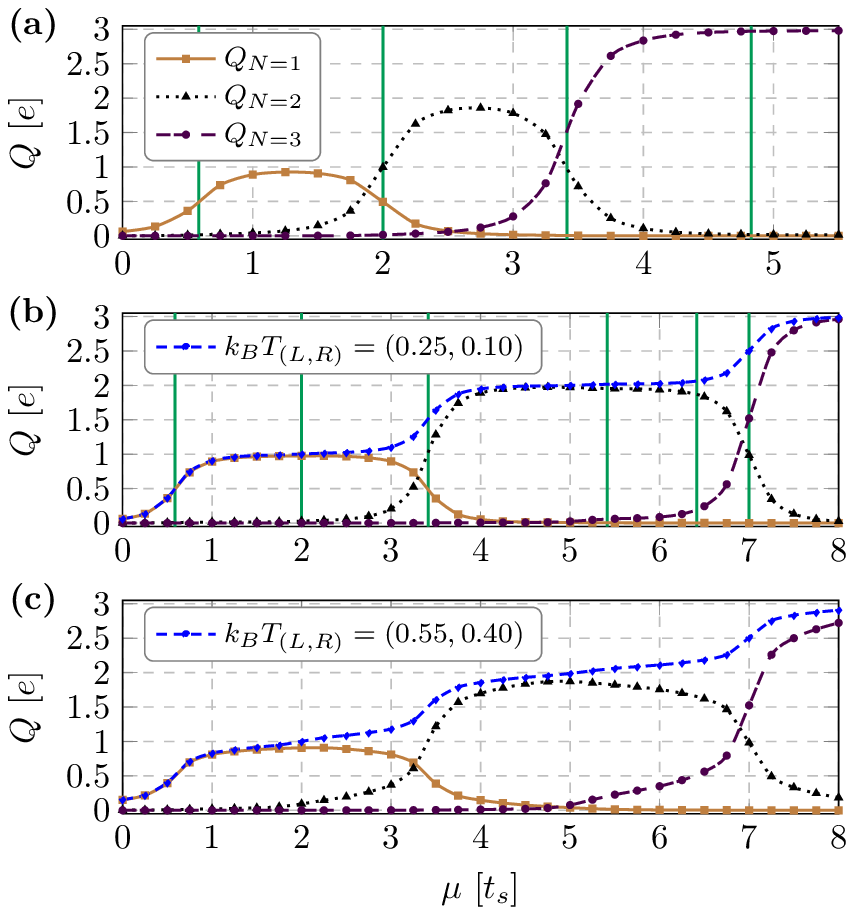}
    \caption{ Charge as a function of the chemical potential.
             Brown solid line~(\brownsolidsquare) shows charging for
             the single particle states, black dotted lines~(\blackdottedtriangle) for two-particle states,
             violet dashed lines~(\violetddashedcircle) for the three-particle state, and blue dashed lines~(\bluedashedcircle)
             show the total charge.
             \textbf{(a)}~Without Coulomb interaction, \({k_BT_L = 0.25}\) and~\({k_BT_L = 0.10}\). 
             \textbf{(b)}~With Coulomb interaction (\(u_c = 2.0\)), \({k_BT_L = 0.25}\) and~\({k_BT_L = 0.10}\).
             \textbf{(c)}~With Coulomb interaction (\(u_c = 2.0\)), \({k_BT_L = 0.55}\) and~\({k_BT_L = 0.40}\).
\label{fig:chg-coul}}
  \end{figure}

Fig.~\ref{fig:coul} shows the current in the left lead in the steady state for both~\({u_c = 0.0}\)
and~\({u_c = 2.0}\).
The \textcolor{ForestGreen}{green} vertical lines represent the chemical potentials in Fig.~\ref{fig:mu-diag}. The blue line~(\bluedashedcircle)
shows the current for a system with no Coulomb interaction. The current is zero at the points~\(\mu \approx 0.5\),~\(\mu = 2\) and~\(\mu \approx 3.5\). These
points correspond to half filling of states. 
In between such points the current is again zero when integer filling occurs (e.~g. for 
$\mu\approx 1.2$).
The filling can be seen in Fig.~\hyperref[fig:chg-coul]{\ref{fig:chg-coul}(a)}, were the charging of the single, two- and three-particle states
is shown for the non-interacting case. At the point~\(\mu \approx 0.5\) the single particle states are half charged, and at~\(\mu = 2\) the two-particle states
are half charged and the single particle half discharged. The last point~\(\mu \approx 3.5\) is the point is where the two-particle states are half discharged
and the three-particle state half filled.
The red line~(\redsolidsquare) shows the current for the same system with Coulomb interaction~(\(u_c = 2.0\)). The effects of the Coulomb is to create
plateaus of zero current at~\({\mu = 2}\) and~\({\mu \approx 5.5}\). These points correspond to integer filling of states.
Fig.~\hyperref[fig:chg-coul]{\ref{fig:chg-coul}(b)} shows the charging for the system with Coulomb interaction (\(u_c = 2.0\)).
The blue line~(\bluedashedcircle) shows the total charge and has integer fillings at \({\mu = 2}\) and \(\mu = 5\).
At \(\mu = 2\) the single particle states~(\brownsolidsquare) are completely filled
and at \(\mu = 5\) the two-particle states~(\blackdottedtriangle) are filled.
While the points \(\mu \approx 0.5\), \(\mu \approx 3.5\) and \(\mu = 7\) correspond to half filling.

\begin{figure}[H]
    \includegraphics[width=1.0\linewidth]{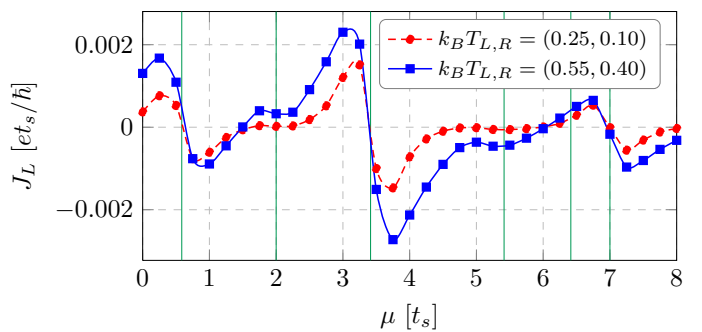}
    \caption{ A comparison of the current in the left lead for two different temperatures.
             The red dashed line~(\reddashedcircle) has~\({k_BT_{L,R} = (0.25, 0.10)}\).
             While the blue solid line~(\bluesolidsquare) has~\({k_BT_{L,R} = (0.55, 0.40)}\).
             The temperature difference between the leads is the same in both cases~\({\Delta k_BT = 0.15}\).
\label{fig:cur-temp}}
  \end{figure}

In Fig.~\ref{fig:cur-temp} we show the current for two different
temperatures. The red dashed line~(\reddashedcircle) has the
temperatures~\({k_BT_L = 0.25}\) and~\({k_BT_L = 0.10}\) in the left and
right leads respectively. The blue solid line~(\bluesolidsquare) has the
same temperature difference between the leads \({\Delta k_BT = 0.15}\)
but the temperature has been increased by~\(0.35\) in both leads. The
temperature in the left lead is then~\({k_BT_L = 0.55}\) and~\({k_BT_R =
0.40}\) in the right lead.  One effect of increasing the temperature in
both leads is that the current increases in magnitude. The zeroes in
the current due to half filling are not affected but the plateaus due
to integer filling are raised or lowered and reduced.  This behavior can
be also explained using the charging diagram. Comparing 
Fig.\ \ref{fig:chg-coul}(c) with Fig.\ \ref{fig:chg-coul}(b) one can see
that around chemical potential $\mu=2$, due to the increased temperature, the
population of single particle states decreases considerably from one, but the 
population of the two-particle states increases considerably from zero. Therefore
the single particle states create a negative current whereas the two-particle states
create a positive one (as illustrated in Fig.\ \ref{fig:num-cur}).  The sum of these
two contributions create the total current, in this case positive.  For chemical 
potentials around 5.5 the total current is a combination of two-particle and 
three-particle components.  The population of the two-particle states is less than 
one, but larger than one half, hence with a negative contribution to the current, 
whereas the three-particle population is between zero and half, hence with a positive 
contribution.  In this case the total current is negative.   

\begin{figure}
    \includegraphics[width=1.0\linewidth]{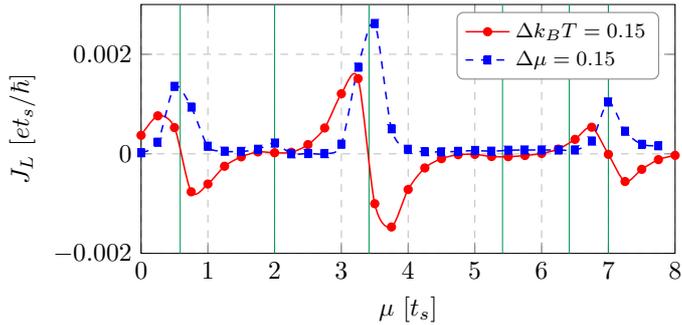}
    \caption{A comparison of the current created by a temperature bias with the current 
created by a chemical potential bias.
             The red solid line~(\redsolidcircle) shows the thermal current with \({k_BT_L = 0.25}\), \({k_BT_R = 0.10}\) vs. the chemical potential $\mu=\mu_L=\mu_R$.
             The blue dashed line~(\bluedashedsquare) shows the electrical current with 
no temperature bias, \({k_BT_L = k_BT_R = 0.10}\), but \({\Delta \mu = \mu_L-\mu_R=0.15}\), i.~e. energetically identical to the the previous thermal bias.
             On the  horizontal axis we use the mean value of the left and right chemical potentials \({\mu = (\mu_L + \mu_R)/2}\).
             The lead-dot coupling parameters are $V_{\ell}=V_L=V_R=0.75$ and the hopping 
energy in the leads is $\tau=4$ (units of $t_s$).}
\label{fig:thermo-electro}
  \end{figure}

It is interesting to compare the current created by a temperature
bias with that due to an electrochemical bias in the many-body case.
This is what we show in Fig.~\ref{fig:thermo-electro}.  The current
driven by the thermal bias is the same as in Fig.~\ref{fig:cur-temp}.
The temperature and chemical potential differences between the left and
right leads are identical on the energy scale. Obviously in the later
case the current is positive, with peaks indicating the half filling
of the ground state with $N$ electrons, and with zero values in between
these peaks, indicating the Coulomb blocking of the transport.  We also
obtain a small peak at $\mu=2$ which corresponds to a small current
going through the first excited single-particle state.  The magnitude
of the Coulomb interaction is incorporated in the energetic separation
between the main peaks and may be estimated from the energetic length
of the zero plateau of the thermoelectric current.

Now we can estimate the level broadening due to the coupling between 
the leads and the dot, $\Gamma$. For example, the first peak of the current
produced by the chemical potential bias around $\mu=0.6$ has a width
at half height of 0.58 (units of $t_s$).   As shown in the previous section
the broadening corresponding to the bias window is $\Delta\mu+2k_BT=0.35$.  The
difference can be attributed to the coupling, i.\ e. $\Gamma=0.58-0.35=0.23$.
This results is consistent with calculations at lower bias windows and 
temperatures (not shown).  In principle the broadening due to coupling
is proportional to the coupling coefficients, Eq.\ \ref{eq:trans-coeff}, 
which are not simple parameters, but a matrix elements and functions of energy
which are hidden in the numerical solution of the GME. 
Nevertheless, in general, from previous studies \cite{1367-2630-11-7-073019, PRBC, Qflute}, 
we expect $\Gamma$ a fraction of $V_{\ell}$, like 25-50\% or so.   
Comparing the estimated $\Gamma=0.23$ with the thermal energies used $0.10<k_BT<0.55$ we can 
conclude that our results correspond to an intermediate parameter regime, in between the situations
presented in Fig.\ \ref{fig:saw-tooth}.

\begin{figure}
    \includegraphics[width=1.0\linewidth]{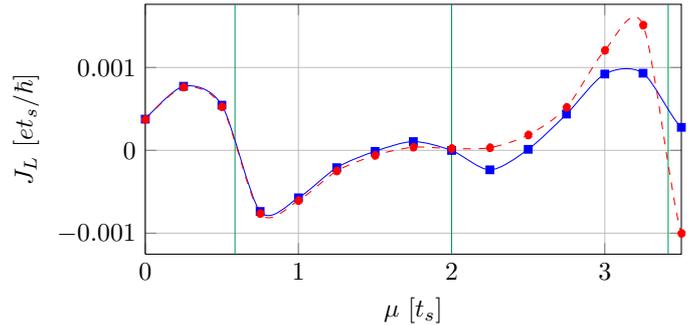}
    \caption{The blue solid line~(\bluesolidsquare) shows the current in the left lead calculated using the GME with \(u_c = 2.0\).
             While the red dashed line~(\reddashedcircle) shows the current calculated with the GME using eigenvectors corresponding to non-interacting electrons, but with
             an energy spectrum forced to be that of the system with \(u_c = 2.0\).}
\label{fig:orthodox}
  \end{figure}

Another comparison we want to make is between the complete Coulomb effects, which 
we describe via the exact diagonalization of the sample Hamiltonian, and the charging
model which is usually invoked in problems related to the Coulomb blockade.  The latter
is known as the ``orthodox" model and it assumes that the energy of a two-particle state
is the sum of the energy of the first two single particle states plus some charging
energy estimated as the electron charge divided by the capacitance of the quantum dot~\cite{Kouwenhoven01}.
This is essentially a mean field assumption.  In order to compare the two approaches 
we calculate the current generated by the temperature bias by using the non-interacting 
eigenvectors in the many-body sample Hamiltonian (i.~e. the basis in the Fock space),
and the ground state energy of the sample for 2 electrons as given by the exact 
treatment.  With this ansatz we describe uncorrelated particles with the energies
of the exact states.  The results are shown in Fig.~\ref{fig:orthodox}.  In the 
``orthodox" model the plateau of the thermoelectric current, at least for a number of 
electrons between one and two, softens or vanishes, possibly depending on strength of the 
Coulomb interaction.  Our interpretation of this result is that the zero plateaus of the 
thermoelectric currents are at least partially and effect of electron-electron correlations, not 
captured by the ``orthodox" model of the Coulomb blockade.

\begin{figure}
    \includegraphics[width=1.0\linewidth]{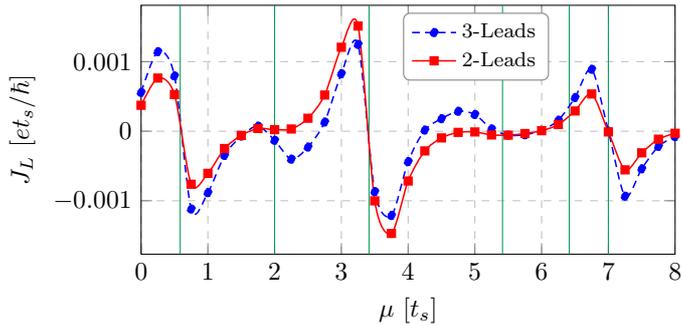}
    \caption{A comparison of the current in the leads between systems with two and three leads.
             The blue dashed line~(\bluedashedcircle) shows the current for a system with two leads, as before, with $k_B T_L=0.25$ and $k_B T_R=0.10$, and the same chemical 
potential $\mu$.
             The red solid line~(\redsolidsquare) shows the current in the same system, but  with a third lead
             attached to the middle point in the sample with $K_B T_L=0.10$ and the 
same chemical potential as the other leads.}
\label{fig:3terminals}
  \end{figure}

Next we show results for a sample attached to three leads, i.~e. with a third lead
connected to site number 2. Such a setup was proposed recently~\cite{Sanchez} in a 
different context, where the third terminal was used as a phase breaking mechanism of the
ballistic electron propagation between the other two terminals. In our case the third
lead has a different effect: it softens the Coulomb blocking of the currents by allowing 
the electrons entering the sample to diffuse into the third terminal. In
Fig.~\ref{fig:3terminals} one can see how the zero plateaus of the thermoelectric current 
vanish in the presence of the third terminal, indicating the additional degree of freedom
offered to the electrons injected into the sample.  

\begin{figure}
    \includegraphics[width=1.0\linewidth]{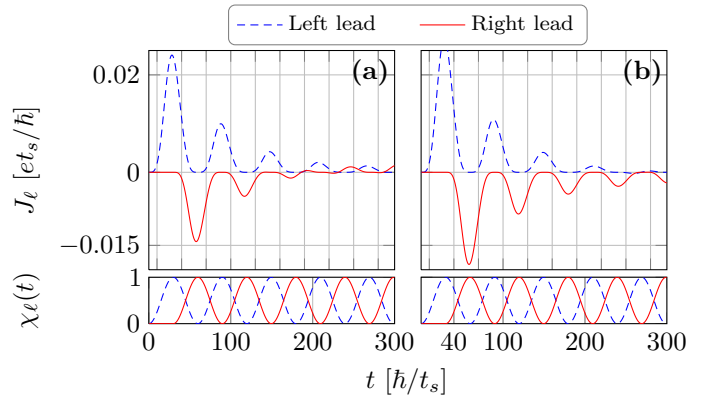}
    \caption{The current in the left lead for a turnstile system. The switching function
             is \({\chi_\ell(t) = \sin (\omega t + \phi_{\ell} )}\) seen at the bottom.
             \textbf{(a)}~Electrochemical bias~\(\mu_{L,R} = 3.25\).
             \textbf{(b)}~Electrochemical bias~\(\mu_{L,R} = 3.75\).
             Other parameters are \({u_c = 2.0}\) and \({k_BT_{(L, R)} = (0.25, 0.10)}\).}
  \label{fig:turnstile}
  \end{figure}

Finally we also show time dependent currents generated by a temperature
bias.  Until now we discussed only the currents in the steady state
obtained by time integration, i.~e. in the asymptotic time limit.
But the important advantage of the GME is that one can examine the
transient regime in the system or another kind of time evolution.
Fig.~\ref{fig:turnstile} shows time dependent currents of a turnstile
model with a thermal bias.  The temperature of the left lead is $k_B T_L =
0.25$ and of the right lead $k_B T_R = 0.10$.  The switching functions
$\chi_{\ell}(t)$ start at zero and continues with sinusoidal pulses,
as shown n the lower panels of Fig.~\ref{fig:turnstile}.  A phase
shift of half a cycle between the left and right lead is included.
The current is in both leads are shown for two cases, $\mu_L=\mu_R=3.25$
and $\mu_L=\mu_R=3.75$.  Earlier, in Eq.~\eqref{eq:JLR}, we used the sign
rule such currents which enter into the sample or exit from it  have
opposite signs.  In this example we prefer to consider positive the
currents from left to right and negative those from right to left.  The
currents follow the shape of the pulses applied to the contacts. Initially
the pulses are large because the system starts out empty. The initial
charging of the system is from both leads, with a positive current from
$L$ into the sample and a negative current from $R$ also into the sample.
After the transient phase the pulses stabilizes and the system enters in
a periodic (steady) state.  In Fig.~\ref{fig:turnstile}(a) the current
pumped in each cycle is positive whereas in Fig.~\ref{fig:turnstile}(b)
the current pumped over each cycle is negative.  This is qualitatively different
from an analog pumping in the presence of chemical potential bias with
the same temperature in the leads.  In that case only positive net current can be
transported over the sample in one complete cycle~\cite{Qflute}.


\section{Concluding remarks\label{sec:conclusion}}

In this work we performed an exploratory study of the current driven
through an open quantum dot in a thermopower setup using the generalized
master equation (GME).  The quantum dot has been connected to two leads having
different temperatures and the same chemical potential.  The quantum dot
has been defined with a finite number of discrete states. The GME method
allows us to describe time dependent, transient, and also steady states,
by numerical integration in time.  We also included the electron-electron
interaction in the sample using the exact diagonalization procedure.
We obtained currents generated by a temperature bias between two leads in
qualitative agreement with the experimental results for the thermopower
of quantum dots at low temperatures~\cite{Svensson}, with saw teeth
alternating with zero plateaus.  The plateaus of the thermopower have
been explained by other authors with the cotunneling processes across the
leads through the dot~\cite{Turek}.  Our formulation of the GME does not
include cotunneling, but only the sequential tunneling.  Therefore in
our model the zero plateaus of the current can be explained using the
level spacing, the Coulomb blocking, and electron-electron correlations.
Cotunneling effects may indeed contribute to the plateaus of the current
and including them into the GME for a multilevel system is a computational 
challenge and the goal of a future work.

Our study may also be compared with a recent work on the thermoelectric
properties of serially coupled quantum dots \cite{Tagani}.  Those
authors used rate equations and a short-range Hubbard model for
the electron-electron interaction and obtained the saw teeth of the
thermopower.  Our sample is designed as a series of sites, and therefore
each site can also be interpreted as a single quantum dot with a single
bound state.  Therefore our model can be naturally used for serially coupled
dots. 

\section{Acknowledgments}
  This work was supported by the Icelandic Research Fund (Rannis). \\

\section{References}

\end{document}